\documentclass[reprint,10pt,nofootinbib]{revtex4-1}

\usepackage[english]{babel}
\usepackage[utf8x]{inputenc}
\usepackage[T1]{fontenc}
\usepackage{multirow}

\usepackage[a4paper,top=3cm,bottom=2cm,left=3cm,right=3cm,marginparwidth=1.75cm]{geometry}

\usepackage{setspace}
\usepackage[]{natbib}

\makeatletter
\renewcommand\@biblabel[1]{}
\makeatother

\title{Community interactions determine role of species in parasite spread amplification: the ecomultiplex network model}

\usepackage{amsmath}
\usepackage{graphicx}
\usepackage[colorinlistoftodos]{todonotes}
\usepackage[colorlinks=true, allcolors=blue]{hyperref}
\usepackage{multirow}

\begin{document}

\title{Community interactions determine role of species in parasite spread amplification: the ecomultiplex network model}

\author{Massimo Stella}
\affiliation{Institute for Complex Systems Simulation, University of Southampton, UK}
\author{Sanja Selakovic}
\affiliation{Faculty of Geosciences, Utrecht University, The Netherlands}
\author{Alberto Antonioni}
\affiliation{Institute for BioComputation and Physics of Complex Systems, University of Zaragoza, Spain}
\affiliation{Grupo Interdisciplinar de Sistemas Complejos, Departamento de Matematicas, Universidad Carlos III de Madrid, Spain}
\author{Cecilia S. Andreazzi}
\affiliation{Campus Fiocruz, Funda{ç}{ã}o Oswaldo Cruz, Brazil}
\email{Corresponding author: candreazzi@fiocruz.br}

\begin{abstract}

Most of zoonoses are multi-host parasites with multiple transmission routes that are usually investigated separately despite their potential interplay. As a unifying framework for modelling parasite spread through different paths of infection, we suggest "ecomultiplex" networks, i.e. multiplex networks representing interacting animal communities with (i) spatial structure and (ii) metabolic scaling. We exploit this ecological framework for testing potential control strategies for \textit{T. cruzii} spread in two real-world ecosystems. Our investigation highlights two interesting results. Firstly, the ecomultiplex topology can be as efficient as more data-demanding epidemiological measures in identifying which species facilitate parasite spread. Secondly, the interplay between predator-prey and host-parasite interactions leads to a phenomenon of parasite amplification in which top predators facilitate \textit{T. cruzii} spread, offering theoretical interpretation of previous empirical findings. Our approach is broadly applicable and could provide novel insights in designing immunisation strategies for pathogens with multiple transmission routes in real-world ecosystems.

\end{abstract}

\maketitle

Zoonoses are infections naturally transmitted between animals and humans, and are the most important cause of emerging and re-emerging diseases in humans \citep{perkins2005role, jones2008global, lloyd2009epidemic}. The majority of the zoonotic agents are multi-host pathogens or parasites \citep{ostfeld2004predators, alexander2012modeling}, whose various host species may differ in their contribution to parasite transmission and persistence over space and time \citep{jansen2015multiple, rushmore2014network}. This heterogeneity of host species contribution to parasite transmission is related to differences in host species' abundance, exposure and susceptibility to infection \citep{haydon2002identifying, altizer2003social, streicker2013differential}. Further, many multi-host parasites have complex life cycles with multiple transmission modes, such as vertical, direct contact, sexual, aerosol, vector-borne and/or food-borne \citep{webster2017acquires}.

Among the zoonotic parasites with multiple hosts and transmission modes, \textit{Trypanosoma cruzi} (Kinetoplastida: Trypanosomatidae), which causes Chagas disease in humans, has complex ecology that challenges transmission modelling and disease control \citep{noireau2009trypanosoma,jansen2015multiple}. \textit{T. cruzi} has already been found in more than 100 mammalian species and its transmission may be mediated by several interdependent mechanisms \citep{noireau2009trypanosoma,jansen2015multiple}. For instance, \textit{T. cruzi} has a contaminative route of transmission that is mediated by invertebrate vectors (Triatominae, eng. kissing bug); and a trophic route of transmission that cascades along the food-web when a susceptible predator feeds on infected prey \citep{noireau2009trypanosoma,jansen2015multiple}. 

Chemical insecticides and housing improvement have been the main strategies for controlling Chagas disease in rural and urban areas of Latin America \citep{dias1999evolution}. However, these strategies are proving to be inefficient \citep{roque2013trypanosoma}. This is possibly related to the maintenance and transmission of parasites among local wild mammalian hosts and its association with sylvatic triatomine vectors \citep{roque2013trypanosoma,roque2008trypanosoma}. Therefore, modelling parasite transmission in a way that is explicitly considering the ecology of wildlife transmission, is fundamental to understanding and predicting outbreaks.

In this work we propose to address this challenge through the mathematical framework of multiplex networks \citep{kivela2014multilayer,boccaletti2014structure,de2016physics, de2013mathematical,battiston2016new}, which are already recognised as a powerful tool in epidemiology \citep{lima2015disease,de2016physics,sanz2014dynamics} and ecology \citep{kefi2015network,kefi2016structured,pilosof2017multilayer,stella2016parasite}. Multiplex networks are multi-layer networks where multi-relational interactions give rise to a collection of network layers so that the same node can engage in different interactions with different neighbours in each layer \citep{kivela2014multilayer,boccaletti2014structure,de2013mathematical}. 

We study the ecology of multi-host parasite spread by multiple routes of transmission and potential control strategies by developing the "ecomultiplex" framework (short for ecological multiplex framework). This framework is innovative because: (i) it accounts for multiple interaction types, reconciling food web structure and parasitic interactions from epidemiological contacts, (ii) it uses metabolic theory \citep{jetz2004scaling} for estimating species frequencies, which are known to influence parasite transmission \citep{mccallum2001should}, and (iii) it considers large-scale realistic spatial structure of wildlife communities \citep{hudson2002ecology}. 

We develop a general model, that could include any ecological interaction among any set of species in real-world ecosystems through ecological multiplex networks. We apply this "ecomultiplex" formalism in investigating parasite spread in two host communities in Brazil: Canastra \citep{rocha2013trypanosoma} and Pantanal \citep{herrera2011food}. We exploit the theoretical framework enriched with empirical data for designing and comparing different wild host immunisation strategies based on: (i) main biological taxonomic groups (e.g. immunising species of a family); (ii) species interaction patterns (e.g. immunising species feeding on the vector); and (iii) species' epidemiological role (e.g. immunising species with higher parasite prevalence). Multiplex network topology proves as powerful as epidemiological field work measurements in predicting the species facilitating parasite spread in both tested ecosystems. More importantly, considering multiple transmission mechanisms confirms the complexity science motto "more is different": on the 2-layers multiplex structure we detect another mechanism for which top predators can indeed facilitate parasite transmission. Our quantitative results challenge the mainstream idea of predators regulating and containing parasite spread in ecosystems \citep{wobeser2013essentials}.

\section{Material and Methods}

\subsection{Ecological multiplex network model}

\begin{figure*}
\centering
\includegraphics[width=0.9\textwidth]{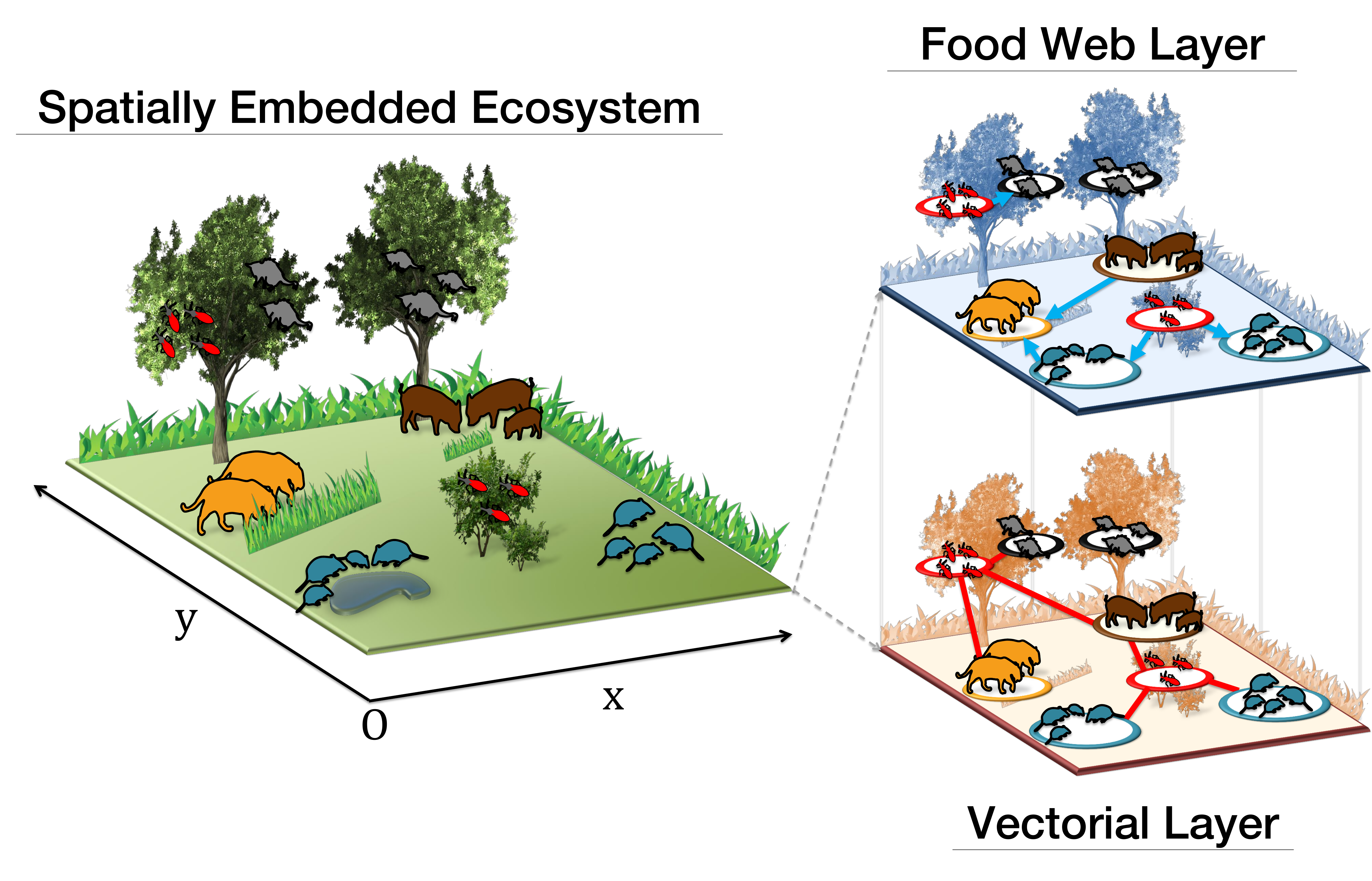}
\caption{\label{intro}Visual representation of our ecological multiplex model. Animal groups are embedded in space and can interact only if they are close enough. Interactions are either predator-prey relationships or host-parasitoid interactions.}
\end{figure*}

The "ecomultiplex" model describes an ecological community interacting in a spatially explicit ecosystem (Fig. \ref{intro}). Each layer of the ecomultiplex represents a different type of interaction between species groups that can potentially lead to parasite transmission. We consider (i) food-web and (ii) contaminative interactions. These interactions give rise to an ecomultiplex network of two layers. Links on the food-web layer are directed to predator species and represent predator-prey interactions. Links on the vectorial layer are undirected and represent vector blood meals of parasitic insects acting as parasite vectors. 
Nodes represent set of individuals from a given species, i.e. animal groups. Distance among animal groups determines possible interactions: only geographically close groups can interact with each other. We fixed the home range of all animal groups as a circle of radius $r=0.03$ over a square of size one and studied a total of $N=10000$ animal groups, cf.~\citet{stella2016parasite}.

\subsection{Ecological data: trophic interactions and body masses}

Predator-prey and vector-host interactions in the ecomultiplex network are based on ecological data related to \textit{T. cruzi} infection in wild hosts within two different areas: Canastra, a tropical savannah in Eastern Brazil \citep{rocha2013trypanosoma} and Pantanal, a vast floodplain in Southern Brazil \citep{herrera2011food}. Both biomes are highly diverse environments where pandemics of \textit{T. cruzi} have been registered \citep{herrera2011food}.

Trophic interactions in the food web are assigned according to literature data about animals’ diets \citep{bueno2002feeding,ramos2007ecologia,cavalcanti2010biologia,amboni2007dieta,dos2012predaccao,reis2006mamiferos, rochaarea} (cf. SI Sect. 1). All vector species are grouped as one functional group due to missing species-level classification. Species prevalence is used to estimate the contaminative interactions in the vectorial layer \citep{rocha2013trypanosoma,herrera2011food}. Positive parasitological diagnostics for \textit{T. cruzi} (hemoculture) are used as a proxy for connections on the vectorial layer, since only individuals with positive parasitaemia (i.e. with high parasite loads in their blood) are able to transmit the parasite \citep{jansen2015multiple}. Body masses of host species represent averages over several available references \citep{herrera2011food,myers2008animal,reis2006mamiferos,bonvicino2008guia,schofield1994triatominae}.

\begin{figure*}
\noindent\fbox{\begin{minipage}{14.8cm}

\subsection{Mathematical formulation for group frequencies}

Geographical proximity and ecological data regulate link creation in the ecomultiplex model. Ecological data, in particular body masses, regulate the frequency of animal groups.

Previous literature \citet{jetz2004scaling} showed that the density $n_i^{-1}$ of individuals of body mass $m_i$ within a home range follows the metabolic scaling:
\begin{equation} \label{eq2}
n_{i}^{-1}=\beta^{-1} R_i^{-1}m_{i}^{3/4}
\end{equation}
when $R_i$ is the species-specific energy supply rate, i.e. the energy resources available to sustain the animal group in a given area and unit of time and $\beta$ a normalisation constant expressing species metabolism. Empirical work has shown that $R_i$ is independent on body mass \citep{jetz2004scaling}. Assuming metabolic theory provides a good approximation for species densities, the above equation can be used for determining the scaling relationship between body mass $m_i$ and frequency $f_i$ of animal groups for species $i$, depending on vector frequency $f_v$ (cf. SI Sect.2):

\begin{equation} \label{eq7}
f_{i}=(1-f_{v})\frac{m_{i}^{-1/4}}{\sum_{j=1}m_{j}^{-1/4}}.
\end{equation}

The above scaling relationship assumes that a fraction of the energy acquired by individual animals gets transformed into body mass at the global population level, i.e. when all individuals of a species are considered. This global constraint leads to frequencies of animal groups scaling as a power-law with exponent $-1/4$, which is different from the coefficient $-3/4$, which always comes from metabolic theory but refers to individuals rather than groups. 

\end{minipage}}
\end{figure*}

\subsection{Metrics for multiplex network analysis}

We investigate the structure of a given ecomultiplex network through the concept of multiplex network cartography \citep{battiston2014structural} (cf. SI Sect. 3). Multiplex cartography provides a map of the centralities of nodes/animal groups in the ecomultiplex network and it is based on two "coordinates". One is the total number of trophic interactions an animal group is involved in (multidegree \cite{de2013mathematical,battiston2014structural}). The higher the multidegree, the more an animal group interacts with other groups. The second coordinate is the ratio of uniform link distribution across layers (participation coefficient \cite{battiston2014structural}), ranging between 0 (when all links of a node are focused in one layer only) and 1 (when all links of a node are uniformly distributed across layers). The higher the participation coefficient the more an animal group will engage in feeding and contaminative interactions in equal measure (see SI for more details).

\subsection{SI Model on the Ecological multiplex Network}

{\renewcommand{\arraystretch}{1.4}%
\begin{table*}[ht!]
\resizebox{\textwidth}{!}{
\begin{tabular}{l p{2.5cm} p{7cm}}
\hline 
Immunisation Type & Strategy Name & Strategy Targets\\
\hline 
\multirow{3}{*}{Ecomultiplex Topological Features} & Insectivores & Species feeding on the vector in a food-web\\
\cline{2-3} 
 & Parasitised Didelphidae & Didelphidae contaminated by the vector on a vectorial layer\tabularnewline
\cline{2-3} 
 & Parasitised Mammals & All species contaminated by the vector on a vectorial layer\tabularnewline
\hline 
\multirow{3}{*}{Biological Features Only} & All Cricetidae & All Cricetidae\tabularnewline
\cline{2-3} 
 & All Didelphidae & All Didelphidae \tabularnewline
\cline{2-3} 
 & Large Mammals & All species with a body mass > 1 kg\tabularnewline
\hline 
\multirow{2}{*}{Epidemiological Features} & Hemoculture N & The $N$ species with the highest likelihood of being found infected with the parasite in field work (see SI).\tabularnewline
\cline{2-3} 
 & Serology N & The $N$ species with the highest likelihood of having been infected with the parasite during their life time (see SI). \tabularnewline
\hline 
\end{tabular}}
\caption{\label{ref:tab} Immunisation types, names and targets of the strategies we tested (cf. SI).}
\end{table*}

Parasite spread is simulated as a Susceptible-Infected (SI) process on the ecomultiplex structure. We assume that parasite transmission among animal groups happens considerably faster than both (i) group creation or extinction and (ii) parasite transmission within groups, so that meta-populations dynamics can be neglected. At each time step, the parasite can spread from an infected group to another one along a connection either in the vectorial (with probability $p$) or food-web (with probability $1-p$) layer. We consider $p$ as a model free parameter called vectorial layer importance, i.e. the rate at which transmission occurs through the consumption of blood by vectors rather than predator-prey feeding interactions. We characterise the SI dynamics at a global scale by defining a \textit{global infection time} $t^*$ as the earliest time at which the parasite reaches its maximum spread within the networked ecosystem \citep{stella2016parasite}.

\subsection{Immunisation Strategies}

Immunisation strategies provide information on how species influence the parasite spread at a global level: immunising species that facilitate parasite spread, the global infection time $t^*$  is expected to increase compared to immunising random species. We focus on immunising only\footnote{By immunising groups at random in ecomultiplex networks with $ N=10000$ nodes, we identified $\phi = 1000$ as the minimum number of groups/nodes that have to be immunised in order to observe increases in infection times compared to the case of random immunisation with a significance level of 5\% (sign test, p-value$<0.01$).} $10\%$ of animal groups in ecomultiplex networks with $10000$ nodes, in either high ($f_v = 0.25$) or low vector frequency scenarios ($f_v = 0.1$). Immunised groups are selected according to three categories of host immunisation strategies focusing on (see also Table 1):

\begin{itemize}
\item \textbf{Biological features}: main taxonomic groups or body mass;
\item \textbf{Ecomultiplex network features}: interaction patterns on the ecomultiplex structure;
\item \textbf{Epidemiological features}: epidemiological measures of parasite prevalence in wildlife.
\end{itemize}

We define the infection time increase $\Delta t_i$ as the normalised difference between the median infection time $t_{s}$ when $\phi=1000$ nodes are immunised according to the strategy $s$ and the median infection time $t_r$ when the same number of nodes is immunised uniformly at random among all mammal groups, $\Delta t_i = \frac{t_i-t_r}{t_r}$. Infection times are averages sampled from 500 simulated replicates. Differences are always tested at 95\% confidence level.

Positive increases imply that the immunisation strategy slowed down the parasite in reaching its maximum spread over the whole ecosystem more than random immunisation. Negative increases imply that random immunisation performs better than the given immunisation strategy in hampering parasite diffusion. 

\section{Results}

Ecomultiplex structure demonstrates high efficiency in designing strategies for slowing down parasite spread in both Canastra and Pantanal. Immunisation experiments also highlight a mechanism where top predators facilitate parasite spread, challenging the mainstream idea of predators containing parasite diffusion \citep{packer2003keeping,hatcher2006parasites,wobeser2013essentials}. 

\subsection{Network Analysis}

Multiplex cartography for both Canastra and Pantanal (Fig. 1) shows that vectors are: (i) more connected and (ii) distribute their links more equally across the ecomultiplex layers than other species. Hence, vectors can get infected in one layer, spread the parasite on another layer with equal likelihood and potentially infect many species: vectors can indeed facilitate parasite spread through their interactions. The local network structure around vectors in Canastra and Pantanal (cf. Fig. 2) shows that vector groups are in the centre of star-like topologies on both network layers. These topological results confirm that \textit{Triatoma} species promote parasite spread. In fact, control strategies for hampering parasite diffusion can focus on vector removal from the environment \citep{yamagata2006control}. However, these strategies are not stable as vector reintroduction can happen shortly after elimination \citep{funk2013identifying}. Hence, we focus on immunisation strategies considering vectors' importance but immunising other species in the ecomultiplex network.

\begin{figure*}
\centering
\includegraphics[width=0.95\textwidth]{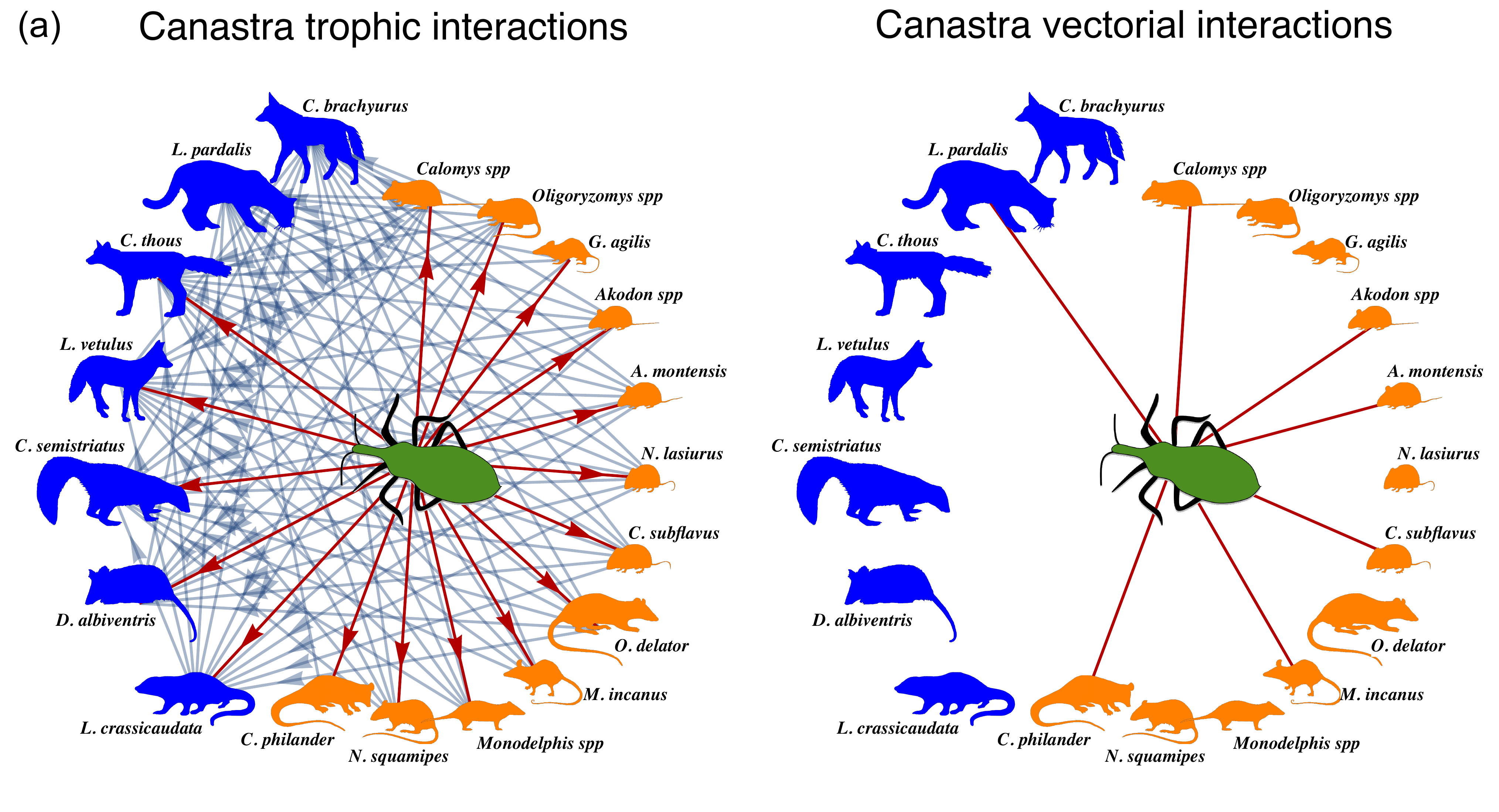}
\includegraphics[width=0.95\textwidth]{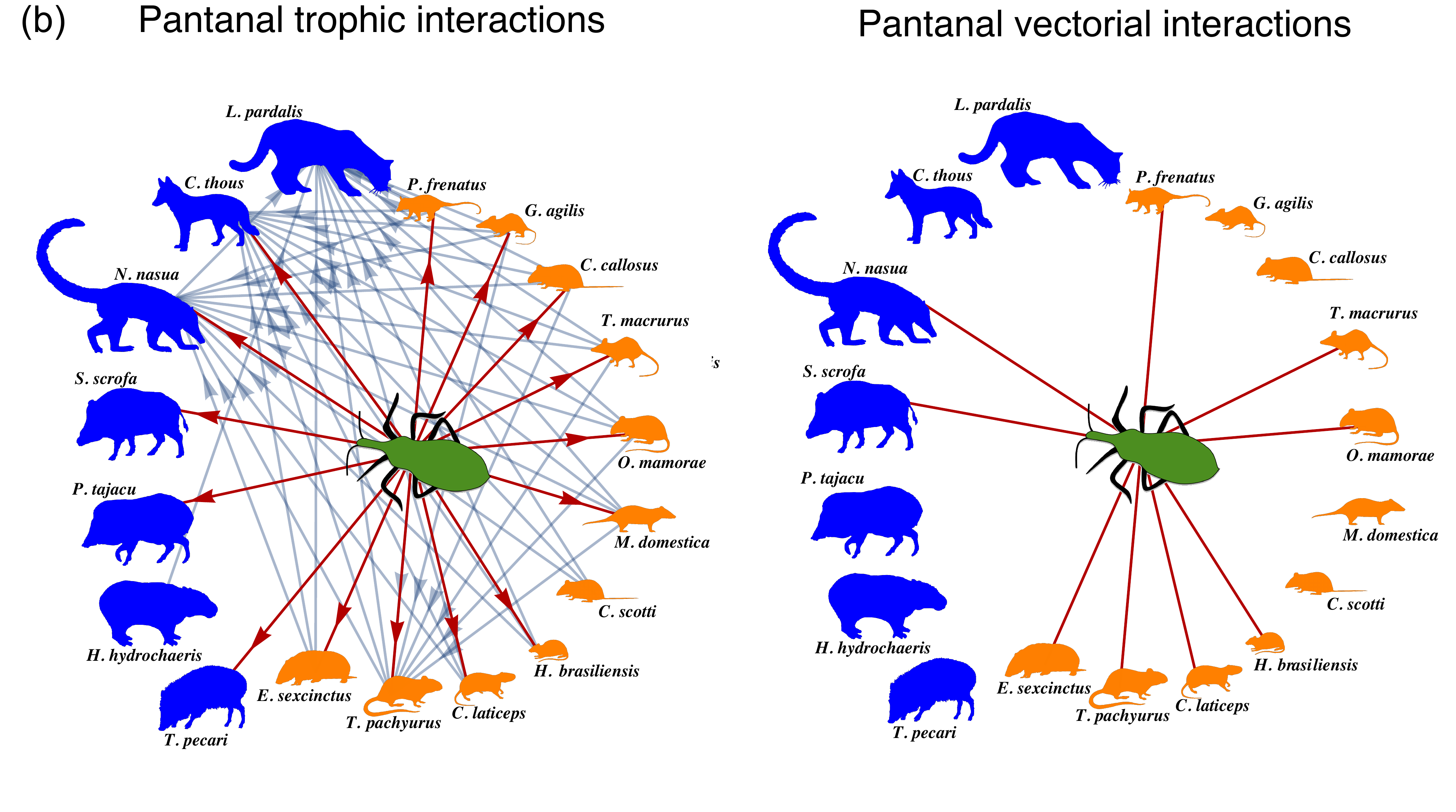}
\caption{\label{fig:visual}Neighbourhood topology of vectors within the Canastra ecosystem on the trophic layer (left) and contaminative layer (right). Predators are highlighted in blue, prey in orange and vectors in green. Interactions involving the insect are highlighted in red. Interactions involving other species are reported for completeness in blue. Vectors are the most highly connected species on the whole multiplex structure: they have the highest outdegree on the trophic layer and the highest degree on the contaminative layer. Furthermore, vectors have the most overlapping connections across the two layers. These findings are reflected in the multiplex cartography.}
\end{figure*}

\begin{figure*}
\centering
\includegraphics[width=0.95\textwidth]{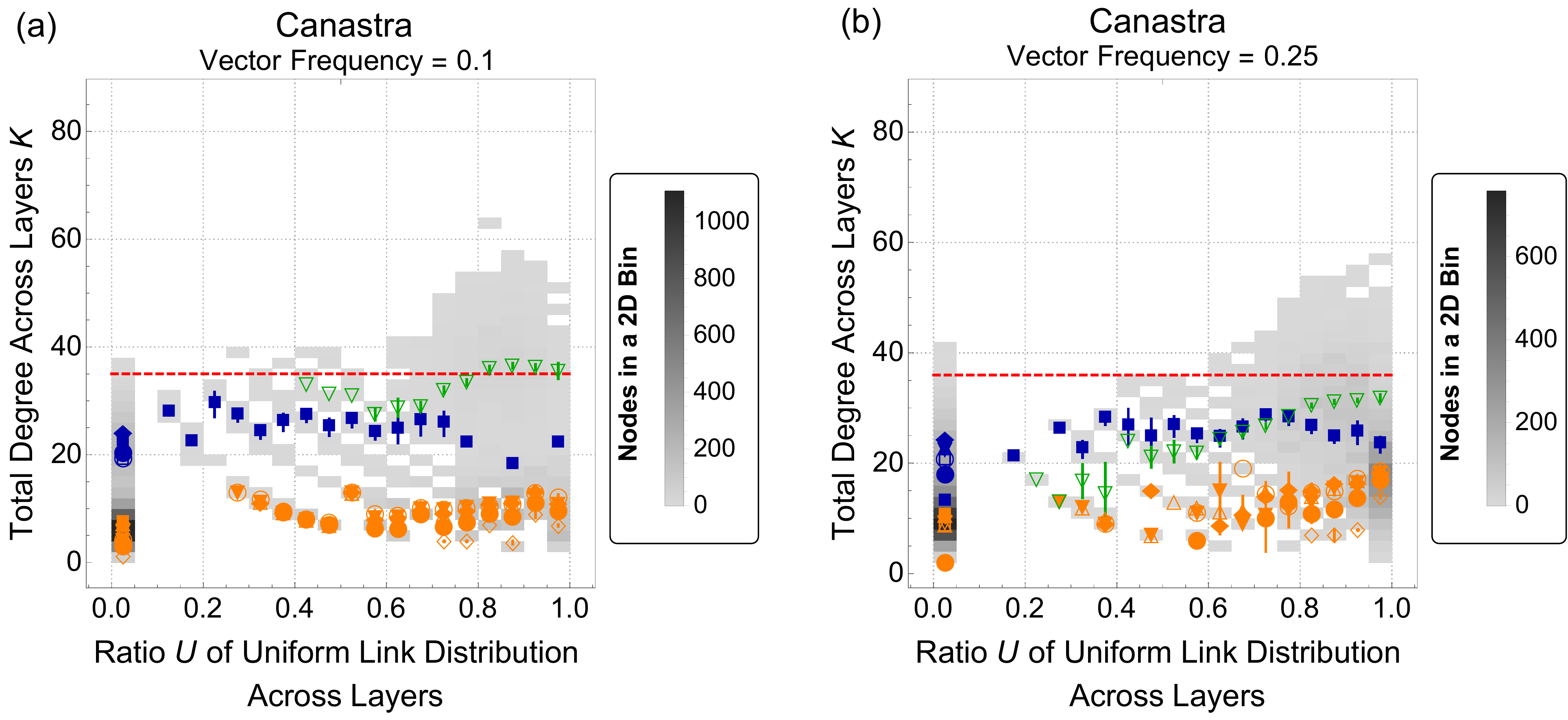}
\includegraphics[width=0.95\textwidth]{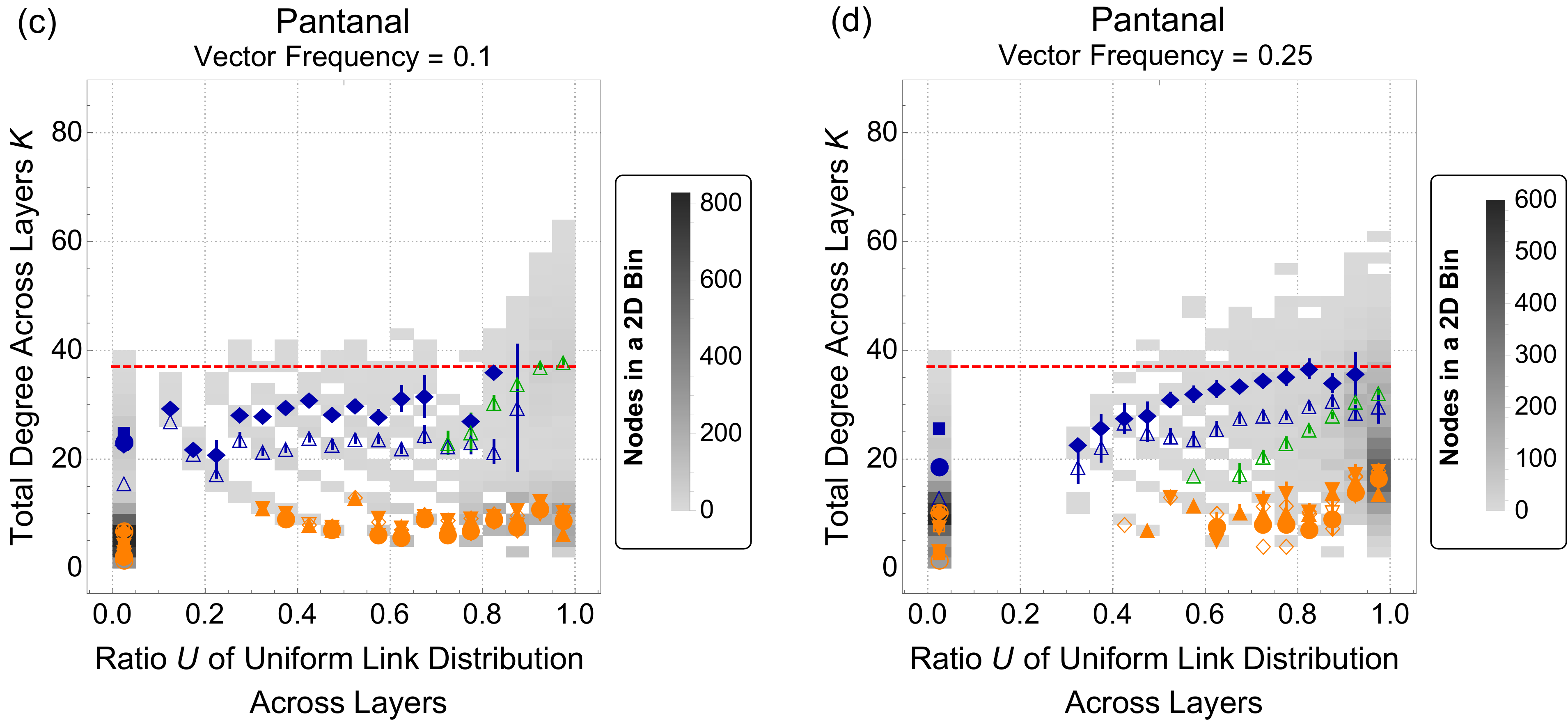}
\caption{\label{fig:carto}Cartography of the ecomultiplex network for the Canastra ecosystem with 10\% (top left) and 25\% (top right) of total groups as vectors. Cartography of the ecomultiplex network for the Pantanal ecosystem with 10\% (bottom left) and 25\% (bottom right) of total groups as vectors. The cartography is presented as a heat map (grey background tiles) and it distinguishes the average trends of species: blue for predators, orange for prey, and green for vectors. Vectors have higher total degree in the ecosystem and tend to distribute more equally their links across both the multiplex layers than all other species. Vectors are therefore pivotal in the ecosystem.}
\end{figure*}

\subsection{Immunisation Strategies}

As expected, immunising species with the highest likelihood of being found infected (an epidemiological strategy) is the best strategy for hampering parasite spread for both Canastra and Pantanal in both vector frequencies scenarios Fig. \ref{fig:immucan}. The epidemiological strategy slows down parasite spread by almost 30\% in Canastra and 26\% in Pantanal when the parasite spreads mainly on the food-web layer ($p_v=0.1$) Fig. \ref{fig:immucan}.  Immunising species interacting with vectors on the vectorial layer (an ecomultiplex strategy) also performs better than random. The difference between the epidemiological and the ecomultiplex strategies is present only at low vector frequencies ($f_v=0.1$) in both Canastra (Fig. 4A) and Pantanal (Fig. 4C) but vanishes when $f_v=0.25$ and $p_v>0.2$ (Fig. 4B,D). 

In Canastra, when 10\% of the animal groups are vector colonies (Fig. 4A), biological immunisation strategies are equivalent to immunising species at random. The performance of biological immunisation changes dramatically when vector colonies become more frequent (Fig. 4B). Immunising large mammals decreases by $12\%$ the global infection time when $p=0.1$: immunising large mammals boosts parasite spread compared to random immunisation. This suggests that large mammals are not the ones facilitating parasite transmission in the model. Immunising all the Didelphidae species leads to similar results (Fig. 4B). Modest increases in infection time are reported for immunising Cricetidae species when $p_v=0.2$ (Fig. 4B). Immunising species feeding on the vector (Insectivores) is equivalent to random immunisation (sign Test, p-values$>0.1$).

In Pantanal, immunising parasitised mammals, parasitised Didelphidae and species with the highest parasite prevalence (hemoculture) are at least two times more effective in slowing down parasite spread compared to other strategies (Fig. 4C-4D). contrary to what happens in Canastra, when $f_v=0.1$ and the parasite spreads mainly on the food web ($p\leq 0.2$), immunising parasitised Didelphidae hampers parasite diffusion more than immunising all parasitised mammals (Sign Test, p-value$<0.01$) (Fig. 4C). Immunising insectivores or large mammals is equivalent to random immunisation (Fig. 4C). Immunising Cricetidae species always performs worse than random immunisation (Fig. 4C,D).

\begin{figure*}
\centering
\includegraphics[height=13.cm]{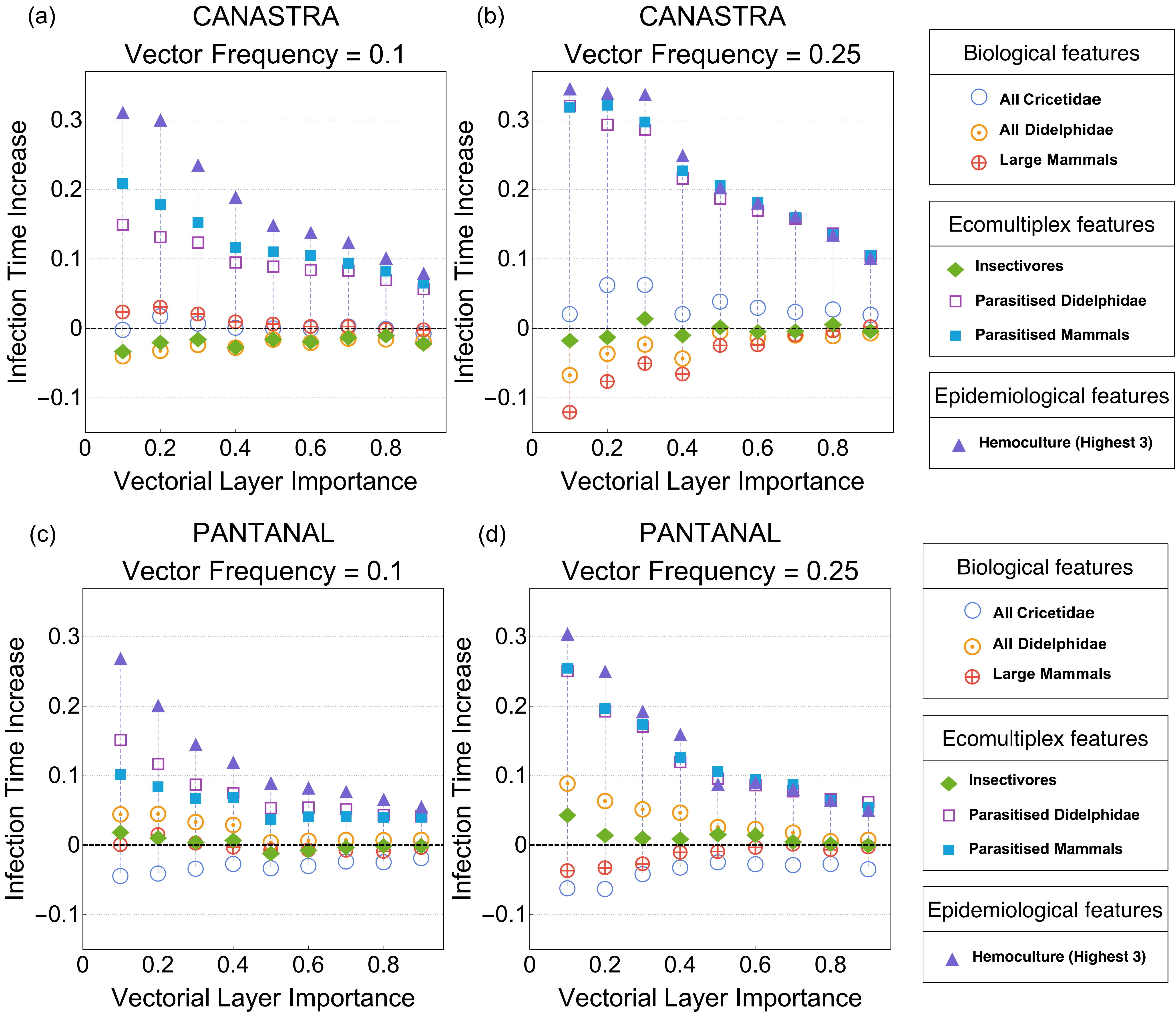}
\caption{\label{fig:immucan}Immunisation strategies for the Canastra (top) and Pantanal (bottom) ecosystems when the vector frequency is 0.1 (left) and 0.25 (right). }
\end{figure*}

\subsection{Top predators can lead to parasite amplification}

In Canastra, the strategy Hemoculture 3 consists of immunising also one species of top predator, the \textit{Leopardus pardalis} (ocelot) (see SI). We compare the performances of Hemoculture 3 against another immunisation strategy where instead of the ocelot we immunise another top predator, the \textit{Chrysocyon brachyurus} (maned wolf), which had negative prevalence in this area \citep{rocha2013trypanosoma}. In general, top predators are related to parasite transmission control in natural environments \cite{wobeser2013essentials} so we do not expect differences.

Instead, results from Fig. 5A indicate a drastic increase of global infection time when a predator with positive parasite prevalence is immunised. This indicates that in Canastra the \textit{Leopardus pardalis} (ocelot) has an amplification effect in spreading the parasite (Fig. 5B). This phenomenon crucially depends on epidemiological importance, as discussed in the following section.

\begin{figure*}
\centering
\includegraphics[width=\textwidth]{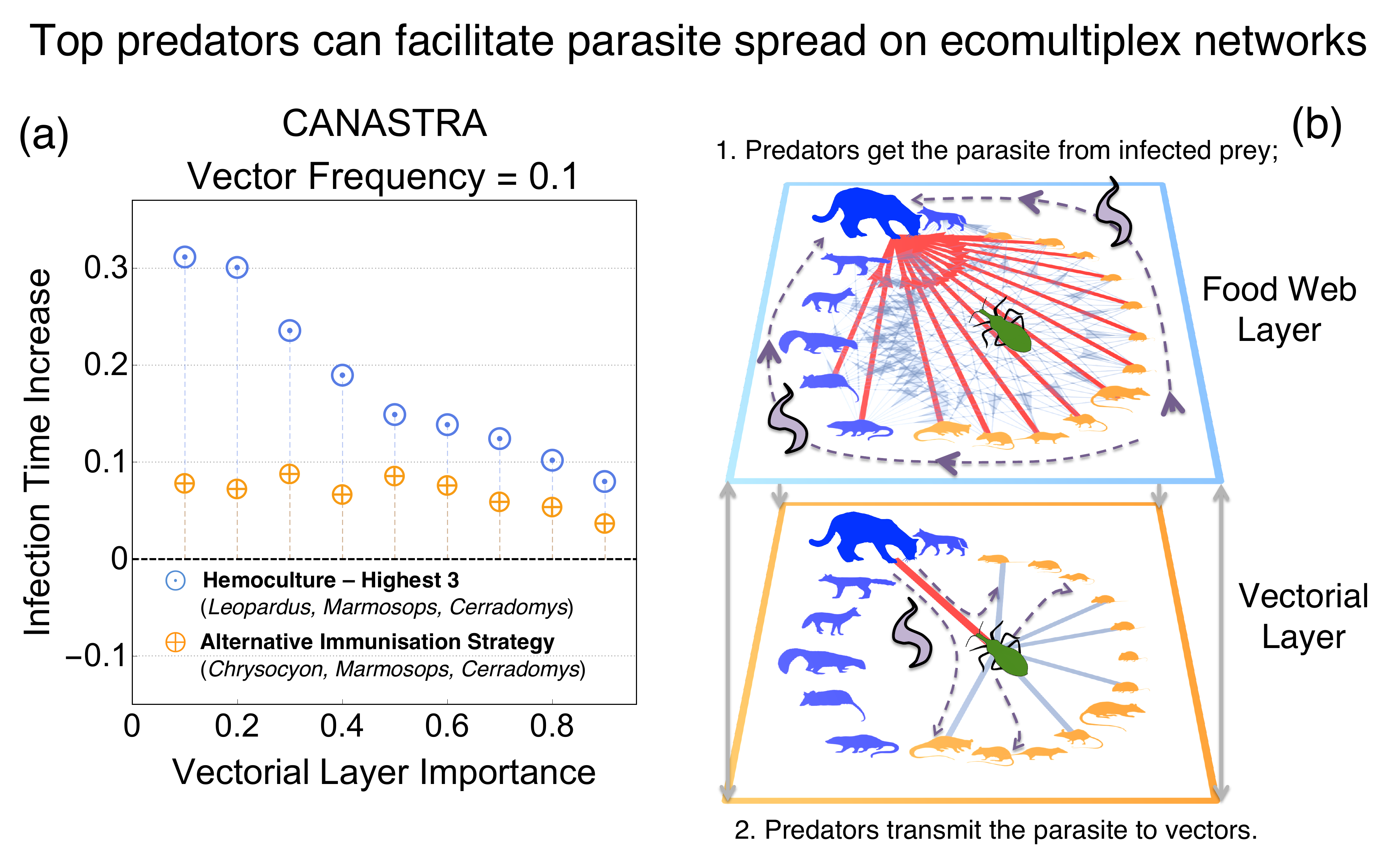}
\caption{\label{fig:leop}Difference in performances of the best immunisation strategy (hemoculture - Highest 3) when instead of the Leopardii the other top predator in the ecosystem (not parasitised by the vector) is immunised instead (hemoculture - H 3 No Leopardus). The other top predator is the maned wolf (Chrysocyon brachiurus).}
\end{figure*}

\section{Discussion}

We present a novel theoretical framework for modelling transmission of multiple-host parasite by multiple routes in real-world ecosystems. We identify three key points related to parasite spread on ecomultiplex networks. First, we show that topological information offers insights on which host species facilitate parasite spread. Second, we use such topological information for designing immunisation strategies for transmission control at different vector abundances. Third, we identify that top predators interaction patterns affect their functional role in parasite transmission, potentially amplifying the parasite spread. 

Network structure is efficient in designing immunisation strategies which perform successfully as epidemiological strategies at higher vector abundances. Ecomultiplex strategies always outperform biological strategies which neglect species' topology. This quantitative evidence suggests the importance of including trophic interactions in case of \textit{T. cruzii} spread \citep{coura2005transmission,johnson2010predators,penczykowski2016understanding}. Although Pantanal and Canastra differ in diversity of species and their interactions, immunising species exposed to parasitic interactions proves efficient in both ecosystems. This underlines the importance of access to the vectorial layer for boosting parasite spread also in the food web. Since the vectorial layer contains only parasitic interactions, our analysis agrees with previous studies \citep{lafferty2008parasites,dunne2013parasites} which underlined the importance of considering the interplay between parasite-host and predator-prey interactions. The ecomultiplex model represents an attempt along the research direction of investigating ecosystems with multiple routes of pathogen transmission \cite{funk2013identifying}.

The ecomultiplex model provides insights on how individual species influence parasite spreading. In Pantanal, immunising only parasitised Didelphidae slows down parasite spread more than immunising all parasitised mammals. This finding is in agreement with previous works identifying Didelphidae as reservoirs for the \textit{T. cruzii} \citep{herrera1992didelphis,noireau2009trypanosoma}, and thus of major importance for facilitating parasite transmission. Notice that our approach identifies Didelphidae as facilitators simply by means of topological interactions, confirming the importance of the ecomultiplex structure in modelling parasite diffusion.

As expected, immunisation strategies based on epidemiological measures are the most effective for control of parasite transmission. However, these strategies require intensive measurements of parasite prevalence across host communities. Instead, ecomultiplex strategies require less data, since building the network requires finding just positive parasitaemia. We show that ecomultiplex strategies slow down parasite spread as much as epidemiological strategies when parasite abundance is high in both Canastra and Pantanal. This shows that ecomultiplex network information is as powerful as epidemiological measurements for gaining insights in the dynamics of parasite diffusion.

Within food webs, top predators are generally considered playing a regulating role in parasite spread by preying on infected individuals and eliminating additional sources of infection for other animals \citep{packer2003keeping,hatcher2006parasites,wobeser2013essentials}. Our ecomultiplex network shows that predators can also facilitate rather than just slow down parasite spread depending on their epidemiological interactions with vectors. An example is the ocelot in Canastra. As reported in Fig. \ref{fig:leop} (b), ocelots are top predators, feed on more prey than other species and have an increased likelihood of becoming infected with the parasite on the food-web layer. Once infected, ocelots can also transmit the parasite to vectors on the vectorial layer. Since vectors themselves facilitate parasite spread, then top predators parasitised by vectors can indeed amplify parasite diffusion. This phenomenon of parasite amplification emerges only when both ecomultiplex layers are considered together. Therefore, this mechanism remarks the importance of unifying ecological and epidemiological approaches for better modelling of multi-host parasite transmission. Interestingly, the amplification mechanism would support previous remarks of ocelot being deeply related with the transmission of \textit{T. cruzii} in wildlife \cite{rocha2013trypanosoma,rochaarea}. 

Our theoretical model allows to design and test immunisation strategies in real-world ecosystems by relying on specific assumptions. For instance, since animal groups are embedded in space, home ranges need to be specified for them. For the sake of simplicity, in this ecological version of the model we considered only one average interaction radius for all species. Considering species-dependent empirical radii (home ranges) represents a challenging yet interesting generalisation for future work. 

Notice that we consider the same parasite transmission probability across species in the SI dynamics. This is because species-dependent transmission rates in our model are encapsulated within: (i) structure of interactions and (ii) different frequencies of animal groups. These two elements play a role equivalent to considering different transmission rates. In previous work \citep{stella2016parasite}, we quantitatively confirmed that considering these two elements in mean-field SI models was sufficient for species to display different probabilities of catching the parasite. Here, immunisation strategies confirm this finding: immunising species that are more exposed to parasites leads to better immunisation performances compared to random immunisation. Considering species-dependent transmission rates as encapsulated in frequencies and network links reduces the number of model parameters.

We assume that parasite spread is happening at much faster rates compared to other meta-population dynamics (e.g extinction or migration), which are not currently considered in the model. However, including meta-population dynamics would allow to explore important research questions such as: (i) the interplay between predation and parasite amplification over top predators influencing parasite spread; (ii) the influence of migration on parasite diffusion; (iii) how extinction patterns influence parasite spread. A promising candidate is the Markovian analytical approach from \citet{gomez2015abrupt}, in order to have even more realistic representations of ecosystems through an ecomultiplex framework.

\section{Acknowledgements}

The authors thank Alireza Goudarzi for insightful discussions and acknowledge the WWCS2017. M.S. was supported by an EPSRC DTC grant (EP/G03690X/1). S.S. acknowledges support from the NWO Complexity grant no. 645.000.013 and ERC Estuaries grant no. 647570. A.A. acknowledges support from the Swiss National Science Foundation under grants no. P2LAP1-161864 and P300P1-171537. C.S.A was supported by Conselho Nacional de Desenvolvimento Cient\'\i fico e Tecnol\'ogico (CNPq/Brazil).

\bibliographystyle{ecol_let}
\bibliography{sample}

\end{document}